# Experimental Observation of Strong Exciton Effects in Graphene Nanoribbons

*Alexander Tries[1,3], Silvio Osella[4], Pengfei Zhang[2], Fugui Xu[2], Charusheela Ramanan[1], Mathias Kläui[3], Yiyong Mai[2], David Beljonne[6] and Hai I. Wang[1,*]*

[1]Max Planck Institute for Polymer Research, Ackermannweg 10, D-55128 Mainz, Germany;

[2] School of Chemistry and Chemical Engineering, Frontiers Science Center for Transformative Molecules, Shanghai Jiao Tong University, 800 Dongchuan Road, Shanghai 200240, China;

[3]Institute of Physics and Graduate School of Excellences Material Science in Mainz, Faculty 08: Physics, Mathematics and Computer Science, Johannes Gutenberg-University Mainz, Staudingerweg 7, D-55128 Mainz, Germany;

[4]Chemical and Biological Systems Simulation Lab, Centre of New Technologies, University of Warsaw, Banacha 2C, 02-097 Warsaw, Poland;

[6]Laboratory for Chemistry of Novel Materials, Université de Mons, Place du Parc, 20, B-7000 Mons, Belgium.

**ABSTRACT**

Graphene nanoribbons (GNRs) with atomically precise width and edge structures are a promising class of nanomaterials for optoelectronics, thanks to their semiconducting nature and high mobility of charge carriers. Understanding the fundamental static optical properties and ultrafast dynamics of charge carrier generation in GNRs is essential for optoelectronic applications. Combining THz spectroscopy and theoretical calculations, we report a strong exciton effect with binding energy up to ∼ 700 meV in liquid-phase-dispersed GNRs with a width of 1.7 nm and an optical bandgap of ~1.6 eV, illustrating the intrinsically strong Coulomb interactions between photogenerated electrons and holes. By tracking the exciton dynamics, we reveal an ultrafast formation of excitons in GNRs with a long lifetime over 100 ps. Our results not only reveal fundamental aspects of excitons in GNRs (gigantic binding energy and ultrafast exciton formation etc.), but also highlight promising properties of GNRs for optoelectronic devices.

**Introduction**

Owing to their massless nature, charge carriers in graphene can possess an extremely high mobility,[1,2] which makes graphene a promising platform for microelectronic[3] and spintronic[4] devices. Yet, its gapless, semi-metallic nature entails severe drawbacks for applications such as electronic transistors and photovoltaics. It has been a long-standing pursuit to open and control the bandgap in graphene, e.g. by tailoring graphene into nanoribbons with atomic precision. Narrow graphene nanoribbons (GNRs) can exhibit a semiconducting behavior with a bandgap due to quantum confinement,[5,6] thus overcoming the lack of usage of graphene in digital logic circuits.[7] By tailoring both the width and the edge structures, GNRs with an optical bandgap of 1-3 eV have been successfully synthesized.[8-17] As the optical absorption of these GNRs lies in the visible and near-infrared part of the electromagnetic spectrum, they are particularly promising for optoelectronic applications including photodetectors and solar cells, by combining the advantages of high charge carrier mobility[13] and tunable light absorption in GNRs.

Thanks to recent bottom up wet-chemistry synthesis breakthroughs,[9] both high quality and sufficient quantity of GNRs are available, facilitating experimental investigation on the optical and optoelectronic properties of GNRs.[18-22] Following previous theoretical studies,[23-28] the electrons and holes generated by optical excitations are subjected to strong Coulomb interactions, resulting in a strongly bounded electron-hole pair, a so-called exciton. The exciton binding energy ($E_B$), defined as the energy required to dissociate the exciton into free electron and hole, has predicted to be substantial, on the order of 1 eV in GNRs, owing to the largely reduced charge screening effects in these atomically flat geometries.[23-28] Consequently, the optical properties of GNRs should be dominated by excitonic resonances, even at room temperature. The nature and dynamics of the excitonic states are of great interest as they are directly related to application-related processes such as light absorption and emission, photoconductivity and electroluminescence and thus motivates several recent studies. For instance, two previous experimental studies on quantifying the exciton binding effect in a seven C-atom wide armchair nanoribbon (7-AGNRs) on gold substrates have been conducted employing reflectance difference spectroscopy[29] and angle-resolved two-photon photoemission spectroscopy.[30] However, due to a polarization effect from the substrate, a modest $E_B$ value of ∼ 160 meV [30] was reported by Bronner et al., much smaller than the theoretical value of 1.8 eV for the same GNRs in the gas phase.[29] Employing transient absorption spectroscopic, Soavi et al.[20] investigated the ultrafast exciton dynamics and unveiled an exciton–exciton annihilation processes in 4CNR-GNRs that possess very large exciton (1.5 eV), and biexciton binding energies (0.25 eV). Huang et al.[31] provided furthermore evidence that the lowest exciton transition in GNR is vibronic in nature. While all these static and ultrafast studies provide strong evidences for a large exciton effect, a direct access and experimental quantification of the exciton binding energy in GNRs remains challenging. Furthermore, little work has been done to elucidate the ultrafast exciton dynamics in GNRS, despite the obvious importance and technological potential of the GNRs.

Previously, different spectroscopic methods have been employed to experimentally quantify $E_B$ in lower dimensional excitonic systems, and each of them has certain advantages and limitations over others. In principle, the binding energy can be inferred directly from absorption or reflection measurements.[32] However, this approach is often limited for nanomaterials with broad exciton transitions due to thermal effects, large size dispersion and their intrinsic interactions with the environment (e.g. lattice vibrations), particularly for systems with small binding energies. In addition, in one-dimensional structures like GNRs, the onset of the single-particle continuum absorption edge is suppressed,[33] further complicating the data interpretation. A way to circumvent this problem is to combine more sophisticated one- and two-photon spectroscopy techniques that quantify the 1S-2P splitting of the exciton states.[34] This approach has been successfully applied for quantifying $E_B$ in varied different nanostructures, including carbon nanotubes[35,36] or conjugated polymers,[37] by further modelling the energy difference between 2P states and the electronic band continuum. However, this method requires a relatively high photoluminescence quantum yield of the materials, which may limit its general applicability. In the recent years, ultrafast THz spectroscopy has been developed to be a powerful complimentary tool to quantify exciton binding energy and further unveil ultrafast formation dynamics of excitons by studying the intra-excitonic resonance (e.g. between 1S-2P transitions). In spite of its great success, due to limited bandwidth of conventional THz spectrometer (up to 2.5 THz for commonly used ZnTe emitter), this method has been used mostly for 2D quantum wells with very small binding energy.[29,38]

In this letter, we report a strong, intrinsic exciton effect with a binding energy up to ~ 700 meV in solution-dispersed GNRs with an uniform width of 1.7 nm and an optical bandgap of ~1.6 eV (GNR-AHM, the GNR structure and its side chain N-n-hexadecylmaleimide (AHM) is shown in Fig. 1a).[31] This is achieved by readily monitoring the free carrier generation dynamics in GNRs following excitation with a short laser pulse (with ~ 40 fs duration): we observe a clear photoconductivity transition from an insulating exciton gas (for $h\upsilon < 2.3$ eV) to a free charge generation (or conductive electron-hole plasma) regime (for $h\upsilon > 2.3$ eV), by simply tuning the photon excitation energy. Furthermore, an exciton binding energy of ~550 meV has been inferred from theoretical calculation, in a good agreement with the experimental results. The exciton states are found to be strongly spatially-confined, indicating their molecule-like, exciton nature. By monitoring the time dependent, frequency resolved conductivity following photoexcitation, we are able to disentangle the contribution of free carriers from excitons to the THz conductivity, and subsequently to track the formation and recombination dynamics of exciton on a picosecond time scale. We find that excitons in GNRs can be formed within 0.8 ps from the initial free charges following a direct photoexcitation of charges into conduction band (by 400 nm in our case), as a direct consequence of a gigantic exciton binding and strong localization of the exciton excitation. Further, the generated excitons are found to be long-lived over 100 ps, rendering GNRs promising for optoelectronic applications. Our results demonstrate not only fundamental aspects of excitons in GNRs but also highlight the great promise of GNRs for optoelectronic devices.

**Results and discussion**

In Fig. 1a, we show a sketch of the molecular structure of the GNRs used in this study, with a uniform width of 1.7 nm and an average length of 11 nm. The ribbons are decorated with pending Diels-Alder cycloadducts of anthracenyl units and N-n-hexadecylmaleimide (AHM) (see synthetic details in ref[31]). The bulky AHM side groups with a size larger than the π-π stacking distance of graphite may effectively prevent aggregation of multiple ribbons, leading to a dispersion of single ribbons in various organic solvents. Previous optical studies using dynamic light scattering, steady-state and TA spectroscopy confirm that the optical features indeed stem from dispersed single ribbons.[31] We further confirm this by performing concentration-dependent THz measurements (Fig.S1), in which we found little changes in the normalized photoconductivity, indicating the negligible role of aggregation in the concentration used in the study. In order to access the intrinsic excitonic properties, a low dielectric solvent (toluene in this study) is intentionally used to reduce the screening effect from the dielectric environment.

The absorption spectrum of the GNR-AHM is shown in Fig. 1b. In contrast to the constant and featureless ~ 2.3% absorption of graphene in the IR and visible range, GNR-AHM show distinct absorption features with two pronounced absorption peaks at 1.63 eV and 1.9 eV, respectively.

Here, we employ optical pump - THz probe (OPTP) spectroscopy to shed light on the nature of the resonances. OPTP has been shown to be a powerful tool for the contact-free characterization of the intrinsic electronic transport properties within isolated GNRs in dispersion.[9,39,40] In a typical OPTP experiment, the GNR-AHM dispersion is photoexcited by ultrashort laser pulses (~ 40 fs) with variable wavelengths (See Fig. 1b). Following such an excitation scheme, charge carriers at different charge states (e.g. the excited states of excitons, or free electrons and holes at valence or conduction band continuum) can be directly populated. Subsequently, the pump-induced photoconductivity is probed by recording the transmission of THz pulses through the dispersion (see supporting information (SI) section 2 for a detailed description). The complex photoconductivity $\Delta\sigma(\omega)$ can then be extracted by comparing the Fourier transform of the THz waveforms transmitted through both pumped- and unpumped samples.[39] The distinction between free carriers and excitons can be readily made, based on the distinct responses in the terahertz frequency range: while free charge carriers are defined by the presence of both real and imaginary conductivity, excitons are manifested by a Lorentz-lineshape resonance (related to the 1S-2P transition[41-44]) in the frequency resolved conductivity. Due to a large exciton binding energy expected in GNRs (>100 THz) according to theoretical results[23,24] and a narrow bandwidth (up to 2 THz) for our THz spectrometer, we expect a pure imaginary component associated with singlet excitons to dominate the dynamics. Therefore, by a time-resolved measurement of the THz spectrum, we can determine the time evolution of excitation dynamics.

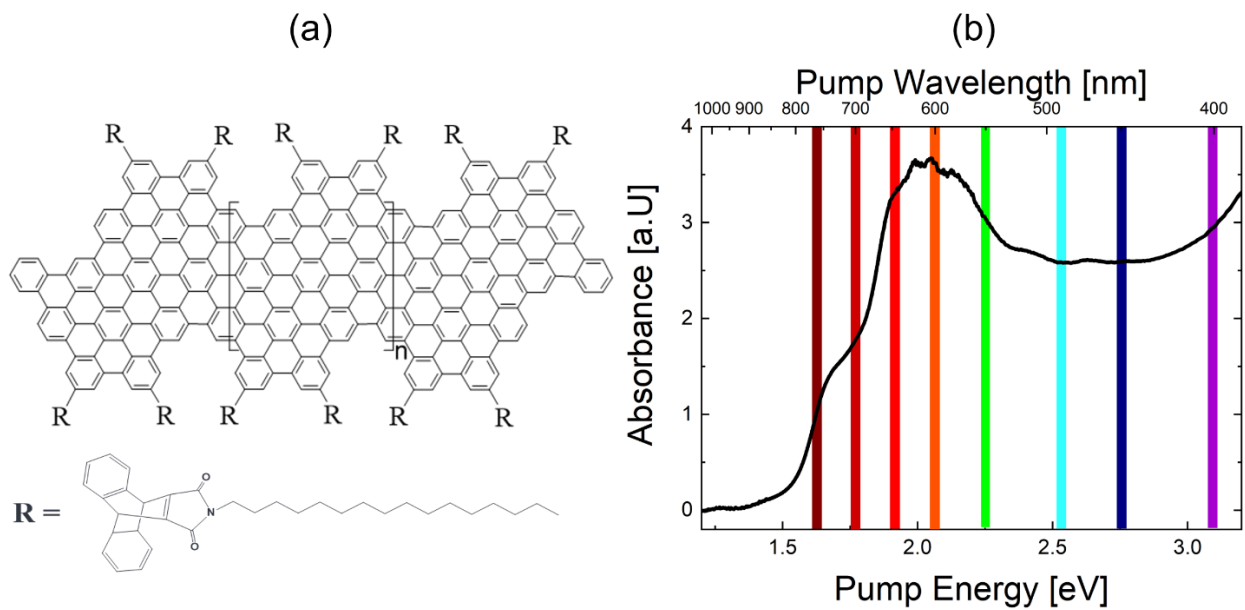

**Figure** 1: (a) Sketch of the GNR-AHM as well as the AHM side group. The GNR-AHM has a repeating unit of 8, thus resulting in an average length of 11 nm with a uniform width of 1.7 nm. The bulky AHM side group has a radius of 0.5 nm. (b) UV/Vis absorption spectrum of GNR-AHM in toluene with a concentration of 1mg/ml. The measurement was corrected for the weak absorption of the solvent. The colored bars represent the pump laser wavelength employed for the optical pump - THz probe measurements to explore the electronic structures of GNR-AHM.

**Charge carrier dynamics in GNRs with different pump energies**. By selectively pumping charge carriers into up to eight different states including the absorption peaks (see Fig. 1b), we monitor and compare the photoinduced optical conductivity $\Delta\sigma(t)$ evolution in the time domain with different excitation energy as shown in Fig. 2a. All data have been normalized to the maximum of the time-dependent imaginary component for a better comparison between the used pump wavelengths. Three dynamics pumped by photon energies of 1.63, 1.9 and 3.1 eV are shown as examples. Taking the concentration $C_{GNR}$ and the molecular weight $M_{GNR}$ into account, the used pump fluences correspond to less than $10^{-5}$ photons per ribbon, thus, we can unambiguously rule out any multi-photon excitation effects that might obscure the desired single-photon dynamics. This is further confirmed by fluence dependent carrier dynamics: as shown in Fig. S2, where we observed a linear increase of the maximum of the real conductivity max ($-\Delta E/E$) with increasing pump laser fluence.

Starting with resonantly pumping the absorption band edge of GNRs (1.63 eV), we observe zero real and a finite imaginary conductivity: a signature of pure exciton states for the first transition at ~ 1.63 eV. By increasing the pump energy further by using 1.9 eV and 3.1 eV excitations, we find in both situations a transient, positive real conductivity of the GNR-AHM with a lifetime of ~200 fs and ~ 700 fs (fittings in Fig. S3a), respectively. The pump energy dependent decay times in the real conductivity are summarized in Fig. S3b, with a ~~fast~~ energy dissipation rate of ~2 eV/1ps (indicating fast exciton formation process in the system as discussed later). Note that such energy dissipation rate is "ultrafast" when compared to the hot carrier decay rates in conventional semiconductors (e.g. GaAs[45] and Ge[46]), and other carbon nanostructures (e.g. graphene[47] and carbon nanotubes[39]), which possess hot carrier lifetimes between 1 to 10 ps. This finite, short-lived real conductivity can be understood as free carrier generation by excess energy assisted exciton dissociation that has been widely reported to be the main free carrier generation mechanism in semiconducting polymers,[43,44,48,49] taking place in an ultrafast sub-100 fs timescale. After the initial ultrafast decay of the real part of the conductivity to nearly zero, the imaginary part is still finite with a similar decay rate as the imaginary part following 1.63 eV pump (at the bandedge). This result suggests that at later times after photoexcitation, the optical conductivity of GNRs is governed by an excitonic response, which is further confirmed by a detailed discussion in the last section of the paper on time-dependent exciton dynamics.

**Experimental quantification of binding energy in GNR-AHM**. First, we quantify the exciton binding energy in the GNR-AHM. The exciton resonance in the frequency domain, as discussed previously, can be described by a Lorentzian resonance originating from 1S-2P excitonic transitions.[41,42] However, owing to the large splitting between 1S-2P intra-excitonic resonances in our system (over 100 THz) and very narrow bandwidth of our THz spectrometer (up to 2 THz), a conventional Lorentzian fitting is not feasible for quantifying a large exciton binding energy $E_B$ in our case (see more discussions in the last section).

Hence, we instead propose a simple, alternative method to determine $E_B$ by tracking the free carrier generation probability in the sample under various pump energies. As discussed earlier in the manuscript, the short-lived real conductivity in Fig. 2a represents the free carrier contribution to the photoinduced THz conductivity. For a quantitative discussion, we plot the peak value of the real part of THz conductivity (normalized to the absorbed photon density $N_{abs}$) versus the pump photon energy $h\upsilon$, as shown in Fig. 2b. We observe a clear exciton to free charge carrier generation transition controlled by the pump energy: at low photon excitation energy (< 2.3 eV), excitation occurs directly into the excitons state manifested by ~ 0 real conductivity; at elevated photon energies (>2.3 eV), free charges are directly formed upon photoexcitation. The essence of this pump energy-dependent (transient) electronic phase control in GNRs can be more quantitatively captured by a simple model depicted in Fig. 2c: the strong Coulomb interaction between the photo-generated electron and hole pair can be described by a deep Coulomb potential with a depth defined by $E_B$. By optically pumping charge carriers into the lowest, 1S state, the probability for exciton dissociation into free charges by thermal fluctuation is nearly zero due to the large Coulomb barrier well in excess of $k_BT$ (thus the observed zero real connectivity by OPTP). With increasing the excess energy $E_{ex}$ for charge carriers (defined as $E_{ex}=h\upsilon- E_{opt}$, with $E_{opt}$ as optical bandgap of GNRs) in the photogenerated electron and hole pairs, the exciton dissociation energy barrier (Δ) is gradually reduced. Here the barrier Δ can be defined as:

$$\Delta = E_B - E_{ex} = (E_B + E_{opt}) - h\upsilon \qquad (1)$$

The reduction in the Δ results in a finite and gradual increase of the free charge generation probability thus increasing the observed real conductance in our THz dynamics. When the excess energy in the photo-generated carriers is sufficiently large to overcome $E_B$, the charge-carrier dynamics are dominant by a free carrier response; and further increasing the pump energy results in a saturation of the real conductivity, as observed in Fig. 2b.

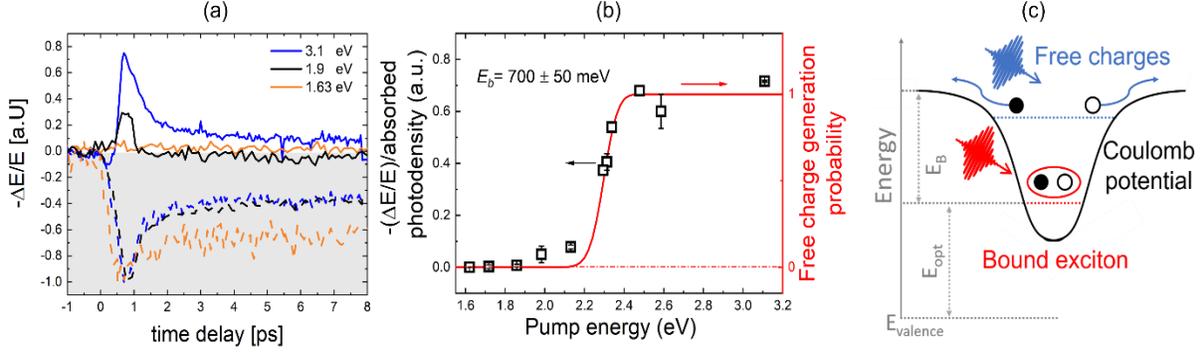

**Figure 2:** (a) The sub-picosecond time evolution of both the real and imaginary frequency-integrated photoconductivity as a function of the pump-probe delay and the pump wavelength. The used fluences were 200 μJ/cm², 246 μJ/cm² and 227 μJ/cm² for 1.63, 1.9 and 3.1 eV respectively; (b) The maximum of the real part of the one-dimensional conductivity normalized to the absorbed photon density. The red line is a best fit to a model described in the main text to account for the free carrier generation probability with increasing the pump energy. The model and the corresponding fitting yield an excitonic binding energy of 700 ± 50 meV; (c) Illustrates the model used here to simulate the probability of exciton dissociation from the deep Coulomb potential into free charges at the band edge by thermal excitation.

In principle, following photoexcitation, free charges and excitons can co-exist in a thermodynamic equilibrium. Besides the aforementioned exciton dissociation via thermal fluctuation over the Coulomb potential, other contributions, such as an entropy-driven dissociation can contribute to the free carrier generation,[38,50] particular in the low excitation density regime. As such, the fraction of excitons is not only governed by the Coulomb interactions between electrons and holes, but also is strongly influenced by the densities of them.[38,50] We argue that such entropic effects contribute very little to the free carrier generation in our study. This statement is supported by the fact that we observed ~0% real conductivity (and thus ~0% free carrier contribution) in the low excitation energy range (1.6-2.2 eV). To further strengthen our discussion here, we applied the thermodynamic Saha equation[38], which reveals 100% exciton species under all the excitation densities employed in our study (see detailed estimation in SI 5). As such, in the following discussion, we will focus on only the effect of thermal excitation on the free carrier generation probability $\eta$ to quantify the exciton binding energy. Here we determine how $\eta$ changes with the energetics of a given state $E$, with the following three considerations:

(i) Thermal fluctuation, $k_BT$ (~ 25.9 meV at room temperature, with $k_B$ the Boltzmann constant and $T$ the temperature) serves as the only driving force for exciton dissociation at a given exciton state $E$ (correspondingly generated by the respective pump wavelength).

(ii) The free carrier generation upon high-energy excitations has been reported to be in the sub-100 fs timescale,[43,44] which is much faster than the time required for cooling and exciton formation processes, both with time scales of hundreds of fs (see the lifetime of the real conductivity for different pump energies in Fig. S3). As such, at the early time scale (such as at the time with the maximum real conductivity), we can neglect the rate competition taking place in the sample, and the free carrier generation probability is solely determined by the thermal excitation as discussed in (i). Combining the discussion of (i) and (ii), the escape probability of electron and hole from the Coulomb potential for a given state E can be expressed as:

$$f(E)= e^{-\frac{E_b+E_{opt}-E}{k_BT}} \text{ (for } E \leq E_b + E_{opt}\text{), or }=100\% \text{ (for } E \geq E_b + E_{opt}\text{)} \quad (2)$$

The latter case describes free carrier generation by pumping carriers directly into the conduction band.

(iii) Lastly, as the excitation is achieved by an ultrafast laser pulse with duration of ~ 40 fs, the energy distribution, or the bandwidth (~ 25 THz, or 100 meV) of the pump pulse becomes very broad, such that we need to take this effect into account. Here, we assume the energy distribution of pump pulses has a Gaussian distribution: $g(E)=\frac{1}{\sqrt{2\pi\sigma^2}}e^{-\frac{E-h\nu}{2\sigma^2}}$, with $h\nu$ as the energy of the selected pump and $\sigma$ the relevant standard deviation ($\sigma$ =12.5 THz =50 meV, the half of the bandwidth). Now taking all assumptions into account, $\eta$ for a selected pump can be simply written as:

$$\eta(h\nu) = \int_0^{+\infty} f(E) * g(E)dE = \int_0^{+\infty} e^{-\frac{E_b+E_{opt}-E}{k_BT}} * \frac{1}{\sqrt{2\pi\sigma^2}}e^{-\frac{E-h\nu}{2\sigma^2}}d(E) \quad (3)$$

Following the very simple model with only two fitting parameters: the exciton binding energy $E_B$ and a normalization prefactor, we can fit the free carrier photoconductivity versus the pump energy very well, as seen in Fig. 2b. Subsequently, from the fitting result $E_B$ can be inferred to be 700 ± 50 meV, indicating intrinsically extremely strong Coulomb interactions between photogenerated electrons and holes. The exciton binding energy for GNR-AHM revealed here, is also in line with previous theoretical predictions of strong exciton effect in a wide range of GNR structures (in the order of ∼ 1 eV).[24]

**Theoretical investigation of strong exciton effects in GNR-AHM**. In order to corroborate our experimental findings of strong excitonic effects for the specific GNR-AHM at hands, and to shed light on the nature of these states, we performed Time-Dependent Density Functional Theory (TD-DFT) calculations on the tetramer (n=2) structure.[51] By resorting to range-separated screened HSE hybrid functional[52] with the 6-31G(d) basis set[53] (as implemented in the Gaussian16 software[54] and described in SI 9, we predict an optical absorption spectrum in Fig. 3 in excellent agreement with the experimental data in Fig. 1b, namely with an intense main optical transition at 1.76 eV and a weaker shoulder at slightly lower energy, ~1.56 eV. A natural transition orbital (NTO) analysis of these transitions reveals that they involve the most frontier molecular orbitals (HOMO and LUMO) at 1.56 eV ($S_0$-$S_1$ transition), and the HOMO-1 and LUMO+1 level at 1.76 eV ($S_0$-$S_5$ transition). To get a deeper insight into the nature of these transitions, attachment and detachment density matrix calculations, probing the electron and hole distribution respectively, were performed. From these, the magnitude of the spatial overlap between the hole and the particle Φs can be evaluated, which directly reflects the nature of the electronic transition (namely the degree of intramolecular charge transfer). The overlaps obtained for GNR-AHM are close to unity and amount to 0.88 for $S_1$ and 0.91 for $S_5$ (Fig. 3), thus clearly pointing to spatially confined excitations, in line with their strong excitonic character inferred from experimental data. We further note that the sharing of the oscillator strength between the two excitons is dictated by the cove-shape topology of the GNR edges, as simple armchairs ribbons feature a single optically allowed transition in this spectral range.[51] Finally, by applying the Polarizable Continuum Model (PCM)[55] and using the electrostatic embedding of toluene as a proxy for environmental effects, we computed the total energies of the neutral and singly charged molecules at the same level of theory. The ionization potential (IP) and electron affinity (EA) obtained as energy differences in such a DeltaSCF approach are IP=4.79 eV and EA=2.68 eV. From the resulting electronic band gap of 2.11 eV, and considering the lowest exciton state at 1.56 eV, the (TD)DFT HSE calculations yield an exciton binding energy of ~550 meV (2.11-1.56 eV), in excellent agreement with the experimental measurement of 700 ± 50 meV. As we are dealing with molecular systems, we explore how the excitonic effects are influenced by the size of the model and, through comparison to experiment, infer an estimate of what the coherence length should be. Calculations on the hexamer yield IP=2.88 eV and EA=4.71 eV for a bandgap of 1.82 eV. This, together with the red-shifted lowest exciton band at 1.46 eV, gives rise to a binding energy of 360 meV for the lowest exciton (Figures S6). These data should be compared to the corresponding values obtained for the tetramer reported above. While the tetramer data qualitatively match the experimental data, the agreement is deteriorating when using a hexamer model instead. We take this as an indication that the effective conjugation length along the ribbons is limited to ~ 4 repeating units (~ 6 nm in length). Even more interesting, our calculations show the presence of two optical absorption bands in the tetramer, with the lowest one involving a longitudinal rearrangement of the electronic density upon excitation, while the second one has a more pronounced transversal character and is in fact reminiscent of the monomer. This is borne out by the fact that the excitation energy for the second optical absorption band remains at ~1.8 eV (calculated value), while the lowest exciton transition shifts from ~1.6 eV to ~1.45 eV when going from the tetramer to the hexamer. Thus, the energy difference between the two excitons can be seen as a measure for the effective conjugation/coherence length along the ribbon. The measured energy spacing between the two exciton bands of ~0.25 eV matches very well that calculated for the tetramer (~0.2 eV) while the corresponding energy difference in the hexamer (~0.35 eV) is slightly larger. This again points to a conjugation length that should not be much larger than 4 monomers. It is worth to commenting that, while the exciton binding energy of our GNRs is in the similar range to two-dimensional transition metal dichalcogenides (TMDC) monolayers such as $MoS_2$, $WS_2$ etc.[32,56,57]), its molecular, localized exciton nature is in a sharp contrast to delocalized Mott-Wannier excitons in TMDCs. From our calculations, we obtain values of the exciton size in the order of 1.7 and 2.2 nm (from TDDFT) and 0.7 nm (from INDO/SCI), which qualitatively compare to the experimental estimation of 0.7 nm from a kinetic energy analysis (see SI section 9 for more details). Thus, the results point towards the formation of strongly bound, Frenkel-like, excitons in the ribbons. This makes GNRs a unique class of excitonic material platform for fundamental study of exciton physics.

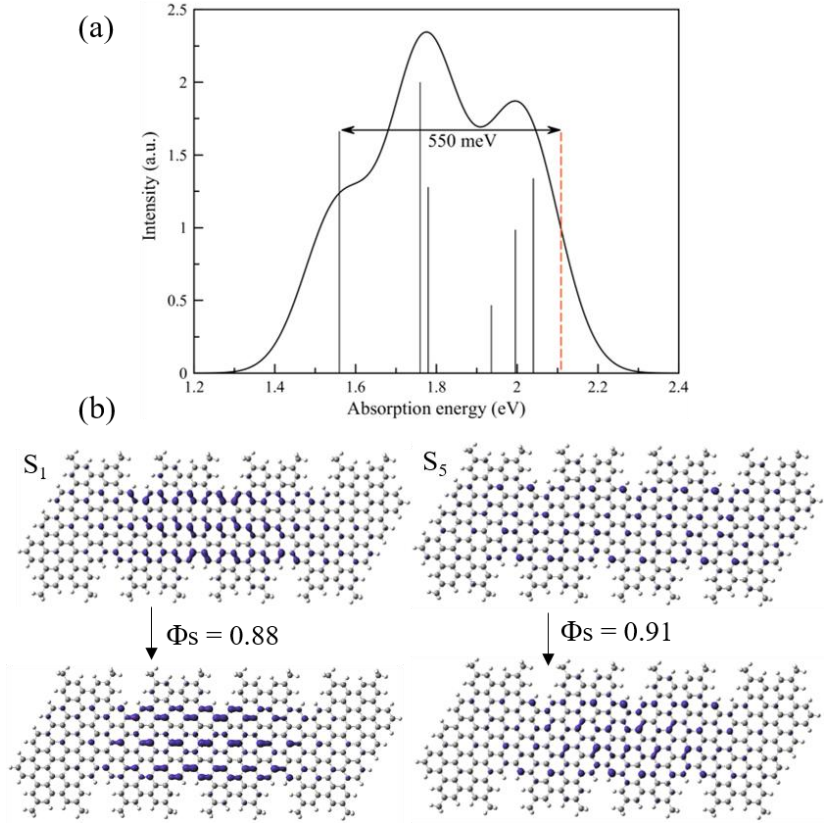

**Figure 3**. (a) Computed absorption spectrum for the GNR-AHM tetramer; black vertical bars represent the oscillator strength for the transitions, the red dashed line the band edge. The binding energy is indicated with a black arrow; (b) detachment/attachment density for the two excited states of interest, S1 and S5, together with their spatial overlap.

**Tracking exciton formation dynamics by non-resonance, high-energy excitations**. Finally, after investigating the gigantic excitonic responses in GNRs, we further track the dynamics of excitons by non-resonance photoexcitation. For that, we measured the delay-time dependent frequency-resolved photoconductivity of our GNRs, following a high-energy excitation by which the charges are optically injected into conduction or valence band continuum (3.1 eV in this case). As this pump energy is much larger than the first absorption transition energy (with an excess energy ~ 1.5 eV), at the early time delay after photoexcitation (e.g. at the maximum of the THz dynamics) the free carrier response is expected to be dominant, following our previous discussion.[40, 58] Indeed, as shown in Fig. 4a, the complex conductivity of GNRs at the maximum value of the photoconductivity is governed by a free carrier response, which can be well fitted by a modified Drude model, the so-called Drude-Smith model (more detailed discussion in SI, section S6). Previously, this charge transport model has been successfully applied to one dimensional carbon nanostructures including GNRs and carbon nanotubes,[9,13,33,40] taking into account the preferential back scattering effect in carrier momentum scattering processes due to structural distortions or the limited length of the structures. Moving away from the photoconductivity peak (e.g. 10 ps after photoexcitation), the complex conductivity is shown to be dominated by an excitonic response as shown in Fig. 4b, identifiable from an almost zero real part conductivity and a negative, increasing imaginary part of the conductivity for higher frequencies. This assignment of an exciton response is further verified by its perfect overlap with the resonance of the exciton state directly generated by band edge excitation (probed at the maximum of the imaginary conductivity upon photoexcitation with 1.63 eV), as shown in Fig. 4b. As such, our result here reveals that photogenerated free charges evolve to reach their thermal equilibrium exciton state in a sub-10 ps time scale following a high-energetic, beyond band gap excitation. In addition, following previous discussion, we attempted to fit the exciton response by a Lorentzian resonance to infer the 1S-2P intra-excitonic transition energy in our systems. As shown in Fig. 4(b) and Fig. S4, several sets of fitting parameters (with varied resonance centers, i.e. 1S-1P transition energies, and broadenings) are shown to fit our data well. In line with our expectation, we can conclude with these fitting that an assessment of the exciton energetics via a conventional Lorentzian fitting over a fairly small frequency bandwidth (up to 2 THz) is not feasible to obtain robust results for GNRs with a large $E_b$ in our case.

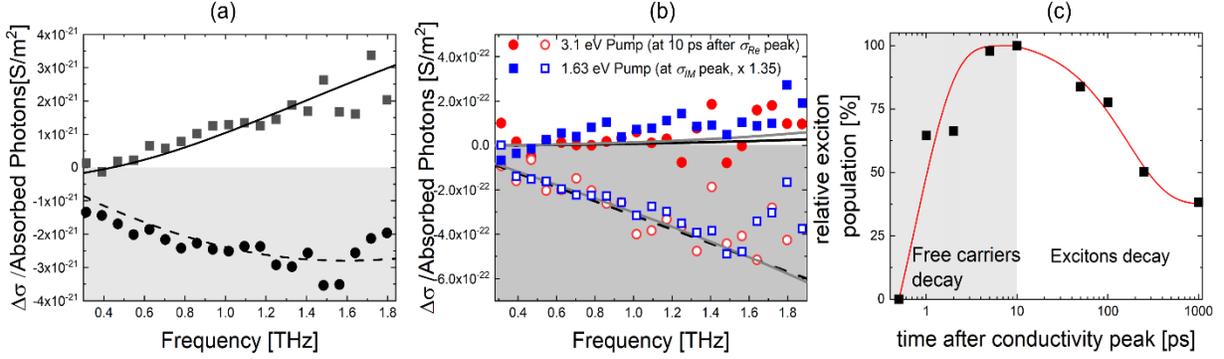

**Figure 4**: (a) The complex frequency-dependent conductivity measured at 0.5 ps, at the peak of the photoconductivity by 3.1 eV pump. The conductivity is scaled to the density *N* of absorbed photons; the solid line represent the Drude-Smith fitting described in the SI; (b) The complex frequency-dependent conductivity comparison between the one measured at 10 ps after photoexcitation with 3.1 eV pump (read), and that at the peak of the imaginary conductivity with 1.63 eV pump (blue, rescale with a factor of 1.35 for comparison). The data are fitted by Lorenztian model described in the main text. Two fitting examples are given here with the center frequency for the black one 25 THz, and the grey one 6 THz. (c) Tracking of exciton formation and recombination dynamics in GNRs, inferred from the fitting described in the main text. The red solid line is a fitting combined an exponential ingrowth and a followed decay, from which the exciton formation and decay time are obtained.

Finally, in order to track the quantitative exciton formation and recombination dynamics, we have fitted frequency resolved dynamics at different time scales. The fitting is done by assuming a simple linear superposition of the free charge response ($\sigma_{\text{free}}$, Drude-Smith-like as shown in Fig. 4a) and pure exciton dynamics ($\sigma_{\text{ex}}$, Lorenztian-type as shown in Fig. 4b):

$$\sigma(t) = \eta_{e-h}(t) * \sigma_{free} + \eta_{ex}(t) * \sigma_{ex}, \qquad (4)$$

with $\eta_{e-h}(t)$ and $\eta_{ex}(t)$ as the contribution from free charges and excitons at a given time. As we can see the fitting examples in Fig. S5, in spite of the simplicity, the model can fit all dynamics well. Based on the fitting, the contribution, or the relative exciton population following photoexcitation has been summarized in Fig. 4c. To this end, based on the fitting, the quantitative exciton formation time can be inferred to be 0.8 ± 0.1 ps, and the following recombination time is found to go beyond 100 ps. This >100 ps exciton lifetime agrees well with two previous transient absorption measurements on GNRs with similar length. In spite of slightly different structures of GNRs, the exciton states are found to be relatively long-lived with lifetimes on the order of 100 ps.[20,31] In addition, the sub-ps exciton formation time reported here indicates an ultrafast energy dissipation rate (~1.9 meV/fs) in the exciton formation process, which is in a sharp contrast to the formation time of several 100 ps in conventional semiconducting quantum wells.[41] Here we briefly discuss the possible origin for the ultrafast energy losses of hot carriers in GNR following an excitation over the band gap. Owing to the soft nature in the structure of GNRs, electron-phonon (*e-ph*) coupling in GNRs could play a critical role on the fast cooling process from hot free carriers into excitons. Indeed, in a recent theoretical study, Zhou et al.[59] provides clear evidence for very strong *e-ph* coupling in GNRs using non-adiabatic molecular dynamics simulations. They found that at room temperature, the GNRs undergo very strong geometrical distortions, which localize the wave functions of charges to enhance both elastic and inelastic *e-ph* scattering in GNRs. This effect leads to accelerated heat dissipation losses, and could explain the fast energy losses during exciton formation in our study. Remarkably, we find a striking agreement between the theoretical calculations and our work in term of the cooling time (~1.9 meV/fs for experimental vs. 1.72 meV/fs for theoretical results). Intriguingly, for the same GNR structure used for this study, a recent experimental work by Huang et al.[31] found the first excited exciton state to be vibronic in nature, indicating the strong interaction between the electronic and vibrational states. A question arises now whether the intrinsic GNR phonons or vibrations from the side chain of the GNR or even solvent molecules dominate the *e-ph* coupling. Given the excellent match between our experimental results and the theory, where the side chains and solvent are not considered, we conclude that the intrinsic phonon modes play a central role on the ultrafast excess energy dissipation in hot carriers. In general, for hot carriers with low electron temperature (i.e. for excitations close to the optical bandgap), previous theoretical studies showed that transverse acoustic (TA) phonons are dominant in the *e-ph* scattering process in GNRs.[60,61] With increasing the excess energy of the photogenerated carriers (and thus the resultant electron temperature of thermalized hot electrons), longitudinal optical (LO) phonon scattering starts to dominate over all other mechanisms.[60] Given that we experimentally deal with a large excess energy, LO phonons of the GNRs are expected to govern the *e-ph* scattering, thus causing a fast energy dissipation of the hot carriers. Further effort (beyond the scope of this work) is required to fundamentally unveil the nature of the *e-ph* coupling in GNRs and to further understand its role on the hot carrier energy dissipation.

Along with fundamental understanding of the exciton dynamics, our experimental results have implications for optoelectronics. Combining a relatively high intrinsic charge mobility in GNRs,[13] the ultrafast formation, and long-lived exciton states make GNRs a promising class of low dimensional for optoelectronic applications, e.g. light emitting diode. This is corroborated by a measurable photoluminescence (PL) of the ribbons in solution: a reasonable quantum yield of 6.3% of the same GNRs is reported.[31] Furthermore, time-dependent PL (Fig. S7) measurements support our THz-measurement for the long lifetime. We find very nice agreement between the exciton decay lifetime by THz and PL, with the latter one yielding measurable PL up to 8 ns (beyond the setup limitations of the OPTP setup). On the other hand, due to the strong electron-hole binding for applications where efficient extraction of charge carriers is required, e.g. photovoltaic, we need to develop an ultrafast sub-ps charge transfer channel to dissociate the transient free charges in GNRs before strongly exciton states are formed.

**Conclusion**

In summary, we report a giant exciton effect with a binding energy up to 700 meV for solution-dispersed GNRs (GNR-AHM in Toluene), illustrating the intrinsically strong Coulomb interactions between photogenerated electron and holes. Following theoretical calculation, we obtain an exciton binding energy of ~550 meV, in excellent agreement with the experimental results; the exciton states are further found to be strongly confined in space, indicating their molecule-like, Frenkel exciton characteristic. As a direct consequence of gigantic exciton binding and strong localization of the exciton excitation, an extremely fast and efficient exciton formation within 0.8 ps in GNRs have been observed. Further, we find that the generated excitons can be long-lived over 100 ps, rendering GNRs promising for optoelectronic applications.

**ASSOCIATED CONTENT**

**Supporting Information**

The Supporting Information is available free of charge on the ACS Publications website.

Concentration-dependent THz measurements, Details on Optical-Pump THz-Probe spectroscopy, fluence-dependent carrier dynamics, details on Drude-Smith fitting, Entropy-driven exciton dissociation, extraction of exciton formation times and fitting examples, results of excitonic fitting with a Lorentzian lineshape, lifetimes of real conductivity, details on computation, Time-dependent photoluminescence.

## AUTHOR INFORMATION

**Corresponding Author**

*E-mail: wanghai@mpip-mainz.mpg.de

**Author Contributions**

H.I.W designed and supervised the project. A.T. carried out all experimental work as well as the fittings. C.R. conducted the time resolved PL measurement. S.O. and D.B. performed the theoretical calculations. P.Z. and F.Z. synthesized the structurally defined GNRs under the direction of Y.M. A.T., S.O. and H.I.W wrote the manuscript. All authors have commented to the manuscript and given approval to the final version of the manuscript.

## ACKNOWLEDGMENT

This work was financially supported by the DFG (Priority Program Graphene SPP 1459, SFB TRR 173 Spin+X) and the Max Planck Society. The authors thank Ivan Ivanov, Xiaoyu Jia, Paniz Soltani and Wenhao Zheng for fruitful discussions, Hansjörg Menges and Walter Scholdei for excellent technical support and Keno Krewer for help with the fitting routines. A.T. is a recipient of a fellowship through the Excellence Initiative by the Graduate School Materials Science in Mainz (GSC 266). Y. M. is grateful for the financial support from the National Natural Science Foundation of China (21774076) and the Program of Shanghai Academic Research Leader (19XD1421700). Y. M. also appreciates the Instrumental Analysis Center at Shanghai Jiao Tong University for some analyses. For the computational time, S.O. thanks the Interdisciplinary Center for Mathematical and Computational Modelling (ICM, University of Warsaw) under the GA-73-16 and GA76-5 computational grants. The work in Mons has been supported by the FNRS-FRFC and Consortium des Equipements de Calcul Intensif - CECI. DB is research director of FNRS.

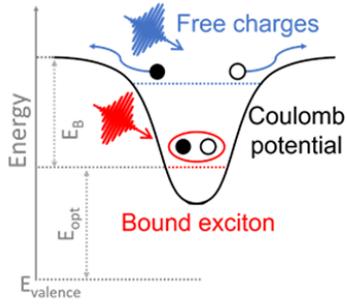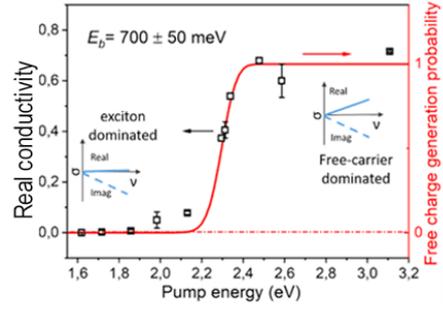

*For TOC only*